\pdfoutput=1
\documentclass[12pt, letterpaper]{article}

\usepackage{ygstyle}

\def\FDR{{\rm FDR}}
\def\FDP{{\rm FDP}}
\def\TPP{{\rm TPP}}
\def\ep{\hfill\blacksquare}
\def\hat{\widehat}

\begin{document}

\title{ADAGES: adaptive aggregation with stability for distributed feature selection}

\author{Yu Gui\thanks{The author finished this paper when he was an undergraduate at the University of Science and Technology of China}}

\maketitle



\begin{abstract}
 In this era of ``big'' data, not only the large amount of data keeps motivating distributed computing, but concerns on data privacy also put forward the emphasis on distributed learning. To conduct feature selection and to control the false discovery rate in a distributed pattern with multi-{\it machines} or multi-{\it institutions}, an efficient aggregation method is necessary. In this paper, we propose an adaptive aggregation method called ADAGES which can be flexibly applied to any machine-wise feature selection method. We will show that our method is capable of controlling the overall FDR with a theoretical foundation while maintaining power as good as the Union aggregation rule in practice.
\end{abstract}

\section{Introduction}
In recent decades, the idea of distributed learning and data decentralization has been frequently discussed. On one hand, the notion of distributed learning is motivated by the advanced techniques of data collection and storage which leads to a large amount of accessible data. Distributed storage and parallel computing are put forward to address the concerns, 
which further requires statistical learning methods in this distributed scenario. On the other hand, statisticians focus on distributed learning since privacy protection is of main interest nowadays. A representative example is the collaborative clinical research among different hospitals on certain diseases, where hospitals will not share patients' data for privacy protection. Therefore, statisticians have to deal with certain ``encoded’’ statistics collected from distributed institutions.

Many recent works focusing on different statistical perspectives have contributed to this field. Estimation is the most fundamental topic in statistics, some works adopt the divide and conquer algorithm for distributed estimation and also study the accuracy of estimation under various contexts, among which are \citep{battey2015distributed}, \citep{JMLR:v16:zhang15d}, \citep{zhao2014general} and \citep{cai2020distributed}. Distributed hypothesis testing is discussed in works such as \citep{ramdas},  \citep{sreekumar2018distributed}, \citep{gilani2019distributed} and is also covered in \citep{battey2015distributed} and \citep{zhao2014general}. Specifically, \citep{su2015communicationefficient}, \citep{Emery2019ControllingTF} and \citep{nguyen2020aggregation} have studied the aggregated feature selection based on multiple knockoffs. Originated from applications, communication constraints and privacy constraints ought to be taken into consideration, \citep{zhangandberger}, \citep{10.1145/2897518.2897582}, \citep{cai2020distributed} study the tradeoff between communication constraints and estimation accuracy. In addition, many other works contribute to distributed learning theories such as \citep{garg2014communication}, \citep{dobriban2018distributed}, \citep{doi:10.1080/01621459.2018.1429274} and \citep{kipnis2019mean}. 
\begin{spacing}{1.5}
\end{spacing}
\noindent\textbf{Controlled feature selection.} 
In addition to feature selection methods such as regularized regression (e.g. \citep{10.2307/2346178},\citep{doi:10.1198/016214501753382273}), controlled feature selection aims to select important features and reduce false selections under some criteria. In this paper, we focus on a fundamental criterion in feature selection: false discovery rate ($\FDR$). The notion of $\FDR$ is introduced in \citep{benjamini1995controlling}. With the definition of the subset $\cS \subset \{1,\dots,d\}$ of relevant features, feature selection is equivalent to recovering $\cS$ based on observations. When the estimated set $\hat{\cS}$ is produced, the false discoveries can be denoted as $\hat{\cS} \cap \cS^{c}$ and {\it false discovery proportion} ($\FDP$) is defined in the form
\begin{align}
{\rm FDP} = \frac{|\hat{\cS} \cap \cS^{c}|}{|\hat{\cS}|}.
\end{align}
The expectation of $\FDP$ is called the {\it false discovery rate} ($\FDR$), i.e. $ {\rm FDR} = \mathbb{E}\left[{\rm FDP}\right]$.
In addition, power of feature selection illustrates the ability to recover true features and thus is defined as
\begin{align}
    {\rm Power} = \EE[\frac{|\hat{\cS} \cap \cS|}{|\cS|}],
\end{align}
which is the expected number of true discoveries over the total number of true features $|\cS |$. 

A series of $\FDR$-based methods originate from the invention of $\FDR$ in \citep{benjamini1995controlling} which utilizes the rank of z-scores for selecting important features. Based on this, \citep{benjamini2001control} relaxes the independence assumption as an extension. Knockoff filter is introduced in \citep{barber2015controlling} with exact control of $\FDR$ and can be extended in a model-free way in \citep{Cands2016PanningFG}. Recently, methods based on mirror statistics are put forward under this topic: \citep{xing2019gm} creates Gaussian mirror variables for all features that get rid of the conditional correlation within each mirrored pair; \citep{dai2020false} utilizes the data splitting and multiple splitting techniques to ensure the recovery of feature importance with stability.
\begin{figure*}[t]
  \centering
  \includegraphics[width=\textwidth]{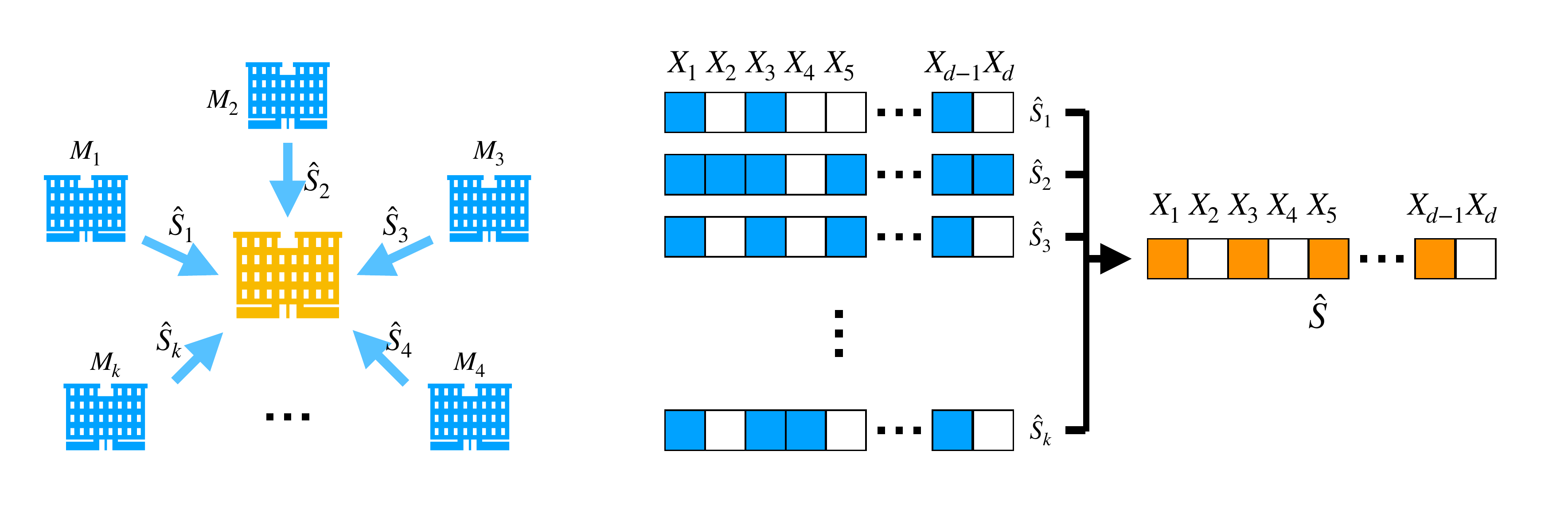}
  \caption{Distributed feature selection and aggregation process}
  \label{fig:plot0}
\end{figure*}
\begin{spacing}{1.5}
\end{spacing}
\noindent\textbf{Stability selection.} As an improvement to general feature selection methods, the notion of stability selection is introduced by \citep{Meinshausen2008StabilityS} which conducts subsampling of size $\left[n/2\right]$ and identifies the most frequently selected features. The idea is close to a ``voting process'' where each sub-sample votes for each feature once and it is in line with our belief that important features will stably become outstanding with more votes. The spirit of stability selection later motivates works such as \citep{Shah2011VariableSW}, \citep{Hofner2015ControllingFD} and also stimulates our idea of adaptive aggregation in distributed feature selection.
\begin{spacing}{1.5}
\end{spacing}
\noindent\textbf{Our contribution.} With the belief in the future of data decentralization, in this paper, we consider the topic of distributed feature selection with a controlled error rate.
We present a general aggregation method for distributed feature selection called ADAGES (\textbf{AD}aptive \textbf{AG}gr\textbf{E}gation with \textbf{S}tability) that can apply for any controlled feature selection method. 
Without looking into the original datasets, we operate on Boolean variables in $\left\{0,1\right\}^{d}$ that is equivalent to subset of features of dimension $d$. Therefore, there is no complex communication or privacy concern in this context. 
Unlike \citep{su2015communicationefficient}, \citep{Emery2019ControllingTF} and \citep{nguyen2020aggregation} that transfer knockoff statistics for aggregation, ADAGES does not depend on any specific feature selection method and is thus more flexible in application. 

Besides, in this paper, we assume the feature selection procedures of all the machines are independent of each other, i.e. as random Boolean vectors, $\hat{\cS}_i \perp \hat{\cS}_j$ for all $i,j \in [k]$. It is noticeable that in practice, the dependence exists due to the overlap of samples for different machines, e.g. the common patients for different hospitals. The generalized case to study the dependence is a promising topic for future work.

\begin{spacing}{1.5}
\end{spacing}
\noindent\textbf{Outline.} We begin with the problem formulation in section~\ref{back} and then in section~\ref{method}, we introduce the detail of ADAGES as an adaptive improvement on empirical rules. In section~\ref{result}, the main theorem will be established to guarantee the exact control of overall \FDR, theoretical proofs of which are in section~\ref{proof}. The results of numerical experiments are shown in section~\ref{simu}.
\begin{spacing}{1.5}
\end{spacing}
\noindent\textbf{Notations.} Suppose the dimension of the $n$ observed features is $d$, i.e. $\bx \in \RR^{d}$. Define $\cS \subset \left\{1,\dots,d\right\}$ is the subset of true features of interest. There are $k$ different machines or institutions contributing to the problem and we denote them as $M_1, \dots, M_k$. For each $i \in \left\{1,\dots,k\right\}$, machine $M_i$ produces an estimated subset $\hat{\cS}_i$ before aggregation and our goal is to obtain $\hat{\cS}$ based on $\left\{\hat{\cS}_1,\dots,\hat{\cS}_k\right\}$. Notation $\hat{\cS}_{(c)}$ refers to the subset produced by the aggregation method with threshold $c$, which will be introduced in section~\ref{method}. Also, $\hat{\cS}_{I}$ and $\hat{\cS}_{U}$ are aggregated subsets of the Intersection rule and the Union rule respectively.

\section{Background}\label{back}
In the context of distributed learning, imagine there is a central machine (the yellow one in Figure \ref{fig:plot0}) and $k$ machines $\left\{M_i:i=1,\dots,k\right\}$ which can be $k$ hospitals or servers. In the current task, the dataset of interest is distributed over all $k$ machines due to concerns of privacy or distance and assume the $i$th machine deals with a sub-dataset $D_i$ with $n_i$ observations. All the machines share the same set of features in the same task, i.e. $\left\{X_j:j=1,\dots,d\right\}$ and they focus on $\FDR$ control with the universal pre-defined level of $q \in \left(0,1\right)$. Suppose the selection result for the $i$th machine is $\hat{\cS}_i$. We should note that the feature selection method adopted for each machine can be arbitrary and the only requirement is that the method should be capable of exact $\FDR$ control. With our adaptive aggregation with stability, we produce the final selection result $\hat{\cS}$ based on controlled selections $\left\{\hat{\cS}_i:i=1,\dots,k\right\}$. For each machine $M_i,i=1,\dots,k$, we define
${\rm FDR}_i = \EE\left[\frac{|\hat{\cS}_i \cap \cS^{c}|}{|\hat{\cS}_i|}\right]$.



\subsection{Empirical aggregation methods for distributed feature selection}
First, three empirical aggregation methods are introduced and we will later cover them as special cases in a generalized family.
Define $z^{\left(i\right)}_j = \bfm{1}_{\left\{j \in \hat{\cS}_i\right\}}$ for each feature, then $\hat{\cS}_i$ is equivalent to an indicator vector $\bz^{\left(i\right)} = \left(z^{\left(i\right)}_1,\dots,z^{\left(i\right)}_d\right)^\top$ and aggregation algorithms can be viewed as operation rules for Boolean variables. Also, in the sense of privacy protection, the selected subset $\hat{\cS}_i$ as the statistics with less sensitive information can be publicly transferred to the ``center machine'' for aggregation. Among aggregation methods, union and intersection of sets are usually adopted empirically.
As the simplest rule similar to the OR rule in Boolean operation, we obtain the Union rule 
\begin{align}
\hat{\cS}_{U} = \bigcup_{i=1}^{k} \hat{\cS}_i.
\end{align}
Also, the intersection of all selected subsets produces the Intersection rule:
\begin{align}
\hat{\cS}_{I} = \bigcap_{i=1}^{k} \hat{\cS}_i.
\end{align}
The Union rule is not strict, thus requires the stricter $\FDR$ control for each machine. It indicates that if each machine has $\FDR$ control at $q$, then the overall $\FDR$ may far exceeds the pre-defined level. On the other hand, the Intersection rule is far more stricter and will result in the loss of power in aggregation. The phenomenon is illustrated in the left plot of Figure \ref{fig:plot1}. We will show that these two rules will have a more general representation and are thus included in a family of threshold-based aggregation rules.

\subsection{Generalized threshold-based aggregation} As an extension to the operation of Boolean variables, we first define 
\begin{align}
    m_j = \sum_{i=1}^{k} \bfm{1}_{\left\{j \in \hat{\cS}_{i}\right\}},\;\;j=1,\dots,d.
\end{align}
Then the {\it threshold-based rule} is conducted as 
\begin{align}
\hat{\cS}_{\left(c\right)} = \left\{j \in \left[d\right]: m_j \geq c\right\}
\end{align}
for an integer $c$.
\begin{rem}
We should notice that {\it the Union rule} is a special case of {\it the threshold-based rule} with $\hat{\cS}_{U} = \hat{\cS}_{\left(c=1\right)}$. And for {\it the Intersection rule},
$\hat{\cS}_{I} = \bigcap_{i=1}^{k} \hat{\cS}_i = \hat{\cS}_{\left(c=k\right)}$.
\end{rem}
Lying between the Intersection and the Union rules, the threshold $c = \left[\left(k+1\right)/2\right]$ can be adopted as a mild rule and we call it ``median-aggregation''. However, we rarely have prior information to determine a universal threshold $c$ and the suitable threshold may also vary in different cases. Therefore, we introduce ADAGES, the adaptive aggregation method in the following section.

\section{Adaptive aggregation for distributed feature selection}\label{method}
Based on the definition of $\hat{\cS}_{\left(c\right)}$, $\hat{\cS}_{\left(c_1\right)} \subseteq \hat{\cS}_{\left(c_2\right)}$ for any $c_1 \geq c_2$, thus $|\hat{\cS}_{\left(c\right)}|$ is a decreasing function of $c$. Further, adaptive information aggregation from $k$ machines utilizes the data-driven threshold which is determined conditionally on $\left\{\hat{\cS}_i,i=1,\dots,k\right\}$, thus it is meaningful to look into the behavior of $\hat{\cS}_{\left(c\right)}|\left(\hat{\cS}_1,\dots,\hat{\cS}_k\right)$. Denote $\bar{s} = \frac{1}{k} \sum_{i=1}^{k} |\hat{\cS}_i|$ and $M = \max_{i=1}^{k} |\hat{\cS}_i|$.

\subsection{Candidate region for threshold}
Restrictions on the size of $\hat{\cS}_{(c)}$ is one traditional way to regularize model complexity, and in the first step, we determine the candidate region for threshold $c$ by restricting the model complexity measure $|\hat{\cS}_{(c)}|$. In the contrast to the usual upper bounds for model complexity, we use the mean $\bar{s} = \frac{1}{k} \sum_{i=1}^{k} |\hat{\cS}_i|$ as a lower bound for $|\hat{\cS}_{\left(c\right)}|$, which is in line with the purpose of power maintenance in multiple testing.

We define $c_0$ as an upper bound as 
\begin{align}\label{c0}
c_0 = \max \left\{c: |\hat{\cS}_{\left(c\right)}| \geq \bar{s}\right\}
\end{align} 
and it is trivial that $c_0 \geq 1$ since 
\begin{align}
    |\hat{\cS}_{U}| = |\hat{\cS}_{\left(c=1\right)}| \geq \max \left\{|\hat{\cS}_i|:i=1,\dots,k\right\} \geq \bar{s}.
\end{align} 
Therefore, we can choose any integer $c \leq c_0$ as a mild threshold for aggregation, but in the meanwhile, a threshold ought to be chosen to balance the tradeoff between false discovery rate and power.

\begin{figure*}[t]
  \centering
  \includegraphics[width=\textwidth]{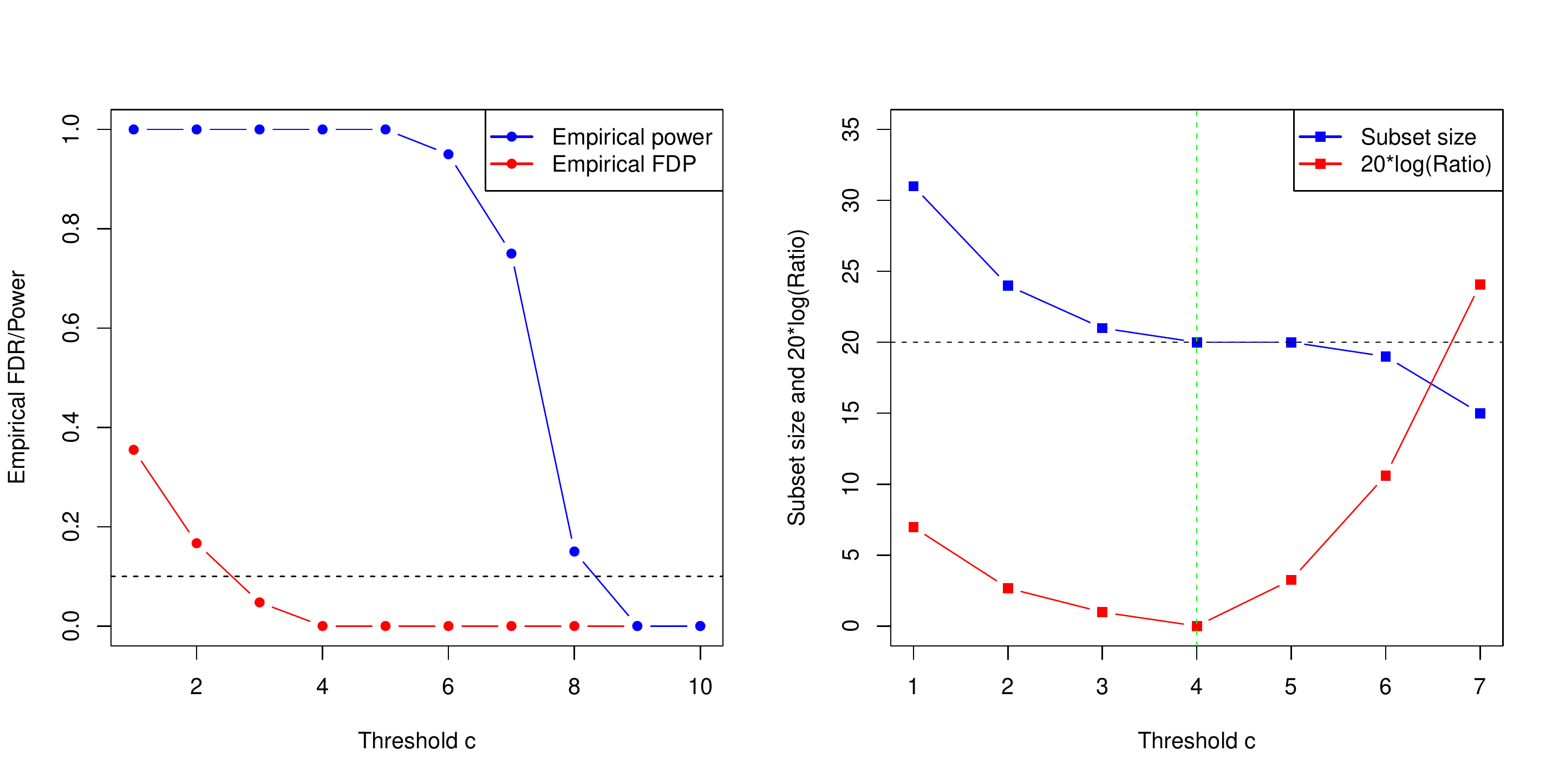}
  \caption{Left: Empirical power and FDP against threshold $c$; Right: Subset size $|\hat{\cS}_{\left(c\right)}|$ and modified-ratio $20*{\rm log}\left(\eta_c\right)$ against threshold $c$. ($n=1000$, $d=50$, $k=10)$}
  \label{fig:plot1}
\end{figure*}
\begin{figure*}[t]
  \centering
  \includegraphics[width=\textwidth]{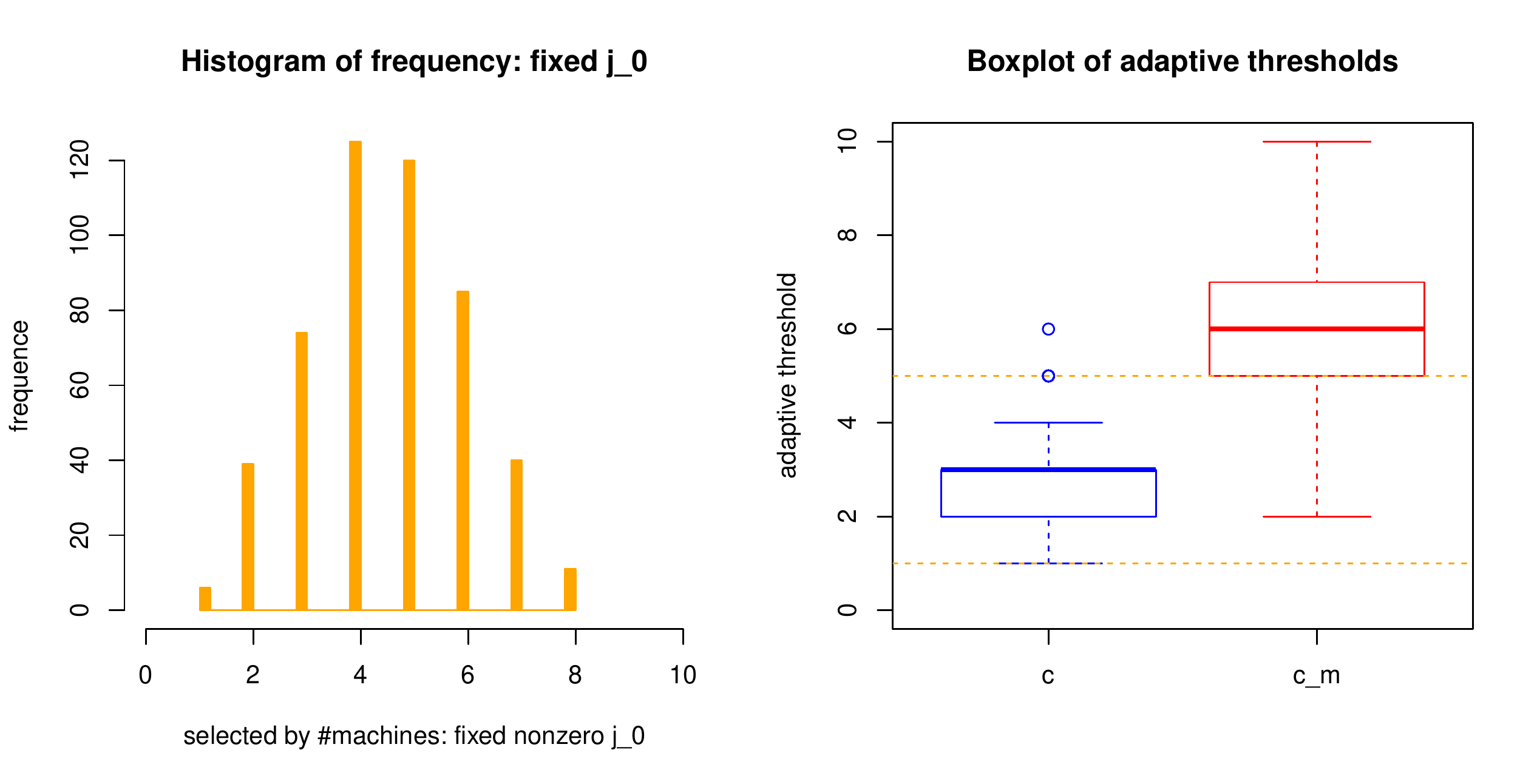}
  \caption{Left: Distribution of appearance in the selected set of $k=10$ machines for a fixed nonzero feature $j_0$ over $500$ trials; Right: Boxplot of adaptive threshold chosen by ADAGES over $500$ trials, $c = c^* = {\rm argmin}_{1 \leq c \leq c_0}\eta_c$ and modified threshold $c_m = \widetilde{c}={\rm argmin}_{1 \leq c \leq c_0} c |\hat{\cS}_{(c)}|$; Dash lines: $c_U = 1$ and $c_M = [(k+1)/2]$~ ($d=50,n=1000,\#nonzeros = 10$).}
  \label{fig:freq}
\end{figure*}
\subsection{Choice of threshold for recovery accuracy}
Besides, to 
improve the tradeoff between $\FDR$ and Power, we adopt the following rule emphasizing stable recovery. With $c_0$ as an upper bound, smaller threshold leads to higher selection power as well as more false discoveries. 
\begin{spacing}{1.5}
\end{spacing}
\noindent\textbf{Complexity ratio.} First, we consider the complexity ratio 
\begin{equation}
\eta_c =
\begin{cases}
\frac{|\hat{\cS}_{\left(c\right)}|}{|\hat{\cS}_{\left(c+1\right)}|},& |\hat{\cS}_{\left(c+1\right)}| > 0,\\
\infty,& |\hat{\cS}_{\left(c+1\right)}| = 0,
\end{cases}
\end{equation}
for thresholds decreasing from $c_0$ and the minimum of complexity ratio is a sign of stable and accurate recovery.
Then, the adaptive threshold $c^*$ for aggregation can be chosen by 
\begin{align}
c^* = {\rm argmin} \left\{\eta_c: 1 \leq c \leq c_0,\right\}.
\end{align}
In practice, to avoid infinite values, we can also use a surrogate $(1+|\hat{\cS}_{\left(c\right)})|/(1+|\hat{\cS}_{\left(c+1\right)}|)$. As is shown in a simple example in the right plot of Figure \ref{fig:plot1} with true $|\cS | = 20$, threshold with the minimum ratio $\eta_c$ produces a more stable recovery of the true $\cS$, and in this figure we adopt a modified form $20 \times {\rm log}\left(\eta_c\right)$ to represent the magnitude of ratio. 
\begin{rem}
To illustrate the complexity ratio, the idea is similar to the eigenvalue ratio in PCA for determining the number of meaningful eigen-components.
We can also consider a toy example where $|\cS \cap \hat{\cS}| \sim B(k,p)$ with $p = \PP(j \in \cS: j \in \hat{\cS}_i)$. In this case, minimizing the ratio $\eta_c$ approximately produces the mode of Bernoulli distribution that  recovers the threshold in line with the most likely frequency for important features.
\end{rem}

\begin{rem}[Threshold-complexity tradeoff]
It is noticeable that another rule with theoretical intuition for choosing the threshold is given by 
\begin{align}
    \widetilde{c} = {\rm argmin}_{1 \leq c \leq c_0} c |\hat{\cS}_{(c)}|,
\end{align}
which explicitly focuses on the tradeoff between the magnitude of threshold and the size of selected subset. As we will show in Lemma~\ref{lem:shrink}, the power shrinkage term $\left(\frac{c|\hat{\cS}_{(c)}|}{k|\cS|} \cdot {\rm FDP}\right)$ plays the leading role in the lower bound for the true positive proportion. Then, for ${\rm FDP}$ at a certain level, minimizing the product $c |\hat{\cS}_{(c)}|$ is equivalent to maximizing the true positive proportion. 
\end{rem}

Details of numerical simulations will be discussed in section~\ref{simu} and the implementation of adaptive aggregation based on the complexity ratio is shown in the Algorithm~\ref{alg1}.
Aggregated feature selection is an initial case dealing with binary variables. It is more exciting to extend this threshold-based aggregation method to estimation and inference based on communication of more informative statistics, and we leave this for future work.

\begin{algorithm}[htb]
\caption{ADAGES: adaptive aggregation with stability for distributed feature selection}\label{alg1}
\begin{algorithmic}[1]
\State \textbf{Input} $\left\{\hat{\cS}_i:i=1,\dots,k\right\}$: $\hat{\cS}_i \subset \left[d\right]$ is the selected subset for the $i$th machine
\State \textbf{Output} $\hat{\cS} = \hat{\cS}_{\left(c^*\right)}$ as an estimation for $\cS$
\State Calculate $m_j = \sum_{i=1}^{k} \bfm{1}_{\left\{j \in \hat{\cS}_{i}\right\}}$, $\forall j \in \left\{1,\dots,d\right\}$
\State Calculate $\bar{s} = \frac{1}{k} \sum_{i=1}^{k} |\hat{\cS}_i|$ 
\For{$c$ in $\left\{1,\dots,k\right\}$}
\State $\hat{\cS}_{\left(c\right)} = \left\{j \in \left[d\right]: m_j \geq c\right\}$
\State Calculate the complexity ratio $\eta_c = \frac{|\hat{\cS}_{\left(c\right)}|+1}{|\hat{\cS}_{\left(c+1\right)}|+1}$, $c \leq k-1$; ~~$\eta_k = \infty$
\EndFor
\State Determine $c_0 = \max \left\{c: |\hat{\cS}_{\left(c\right)}| \geq \bar{s}\right\}$ 
\State Produce adaptive threshold $c^* = {\rm argmin} \left\{\eta_c: 1 \leq c \leq c_0\right\}$\\
\Return $\hat{\cS} = \hat{\cS}_{(c^*)} = \left\{j \in \left[d\right]: m_j \geq c^*\right\}$
\end{algorithmic}
\end{algorithm}

\section{Main result}\label{result}
In this section, we will show the theoretical properties of ADAGES for adaptive aggregation in the scenario of distributed feature selection. First, we obtain the control of overall false discovery rate in theorem~\ref{thm1}; besides, we establish the connection of overall power and machine-wise power: theorem~\ref{thm2} shows the simultaneous control of $\FDR$ and a power shrinkage term and theorem~\ref{thm3} compares the power of ADAGES with the ``optimal'' power produced by the Union rule.

\subsection{Distributed FDR control}
Based on the adaptive threshold for aggregation, the ADAGES produces exact control of the false discovery rate.
\begin{thm}\label{thm1}
For a pre-defined level $q \in \left(0,1\right)$, suppose machine-wise ${\rm FDR}_i \leq q$ for $i=1,\dots,k$ and $\lambda \geq \max_{1\leq i \leq k}\frac{|\hat{\cS}_i|}{c^*} \sum_{j=1}^{k} \frac{1}{|\hat{\cS}_j|}$. Then, ADAGES with $c^* \in \left[1,c_0\right] \cap \ZZ$ produces
\begin{align}
{\rm FDR}_{\left(c^*\right)} = \EE\left[\frac{|\hat{\cS}_{\left(c^*\right)} \cap \cS^{c}|}{|\hat{\cS}_{\left(c^*\right)}|}\right] \leq \lambda q.
\end{align}
\end{thm}

Then, we discuss two special cases with fixed thresholds $c=1$ and $c=k$ respectively, which may reveal their shortcomings to some extend.
\begin{prop}[the Union rule]\label{union}
For a pre-defined level $q \in \left(0,1\right)$, if machine-wise ${\rm FDR}_i \leq q$ for all $i=1,\dots,k$, the Union rule produces
\begin{align}
{\rm FDR}_{U} = \EE\left[\frac{|\hat{\cS}_{U} \cap \cS^{c}|}{|\hat{\cS}_{U}|}\right] \leq kq.
\end{align}
\end{prop}
More generally, as is pointed out in \citep{xie2019aggregated}, if there is a sequence of pre-defined FDR levels $\left(q_1,\dots,q_k\right)$ such that ${\rm FDR}_i \leq q_i$ for all $i \in \left[k\right]$, then the overall $\FDR$ can be exactly controlled at level $q=\sum_{i=1}^{k} q_i$. If we would like to have overall FDR controlled at level $q$, it requires that $\sum_{i=1}^{k} q_i = q$ and a simple case is $q_i = q/k$ for all $k$ machines. Besides, in the case with $c = k$, based on $k|\hat{\cS}_{I} \cap \cS^{c}| \leq \sum_{i=1}^{k} |\hat{\cS}_{i} \cap \cS^{c}|$, 
we have the following proposition:
\begin{prop}[the Intersection rule]\label{intersec}
For a pre-defined level $q \in \left(0,1\right)$, if machine-wise ${\rm FDR}_i \leq q$ for $i=1,\dots,k$ and there is a constant $\kappa \geq 1$
such that
$\max_{i \in \left[k\right]} \frac{|\hat{\cS}_i|}{|\hat{\cS}_{I}|} \leq \kappa$,
then the Intersection rule produces
\begin{align}
{\rm FDR}_{I} = \EE\left[\frac{|\hat{\cS}_{I} \cap \cS^{c}|}{|\hat{\cS}_{I}|}\right] \leq \kappa q.
\end{align}
\end{prop}

Comparing the overall $\FDR$ bounds, the Union rule as a less strict aggregation rule produces $\FDR$ at an expected level as high as $kq$. Instead, the Intersection rule is the most conservative and has theoretical $\FDR$ control at $q$ multiplied by a factor $\kappa$. However, with an adaptive threshold, ADAGES summarizes machine-wise information more efficiently and has the control of overall $\FDR$ at level $\lambda q$. Here, as an illustration, we compare the magnitude of $k,\lambda,\kappa$ to show the abilities of $\FDR$ control of the three methods. First, if $c^*/k$ has a positive lower bound such that $c^* \geq b\cdot k$ and $\max_{i \in [k]} \frac{|\hat{\cS}_i|}{|\hat{\cS}_j|} = O(1)$ for all $j$, then we obtain $\lambda = o(k)$. Comparison between $\lambda$ and $\kappa$ is of more interest, which is summarized in the following proposition. 
\begin{prop}
Denote the tight bound 
\noindent $\bar{\lambda} = \max_{i \in [k]}\frac{|\hat{\cS}_i|}{c^*} \sum_{j=1}^{k} \frac{1}{|\hat{\cS}_j|}$ and $\bar{\kappa} = \max_{i \in \left[k\right]} \frac{|\hat{\cS}_i|}{|\hat{\cS}_{I}|}$. 
Then, we have
\begin{align}
    \frac{\bar{\lambda}}{\bar{\kappa}} = \frac{1}{c^*} \sum_{j=1}^{k} \frac{|\hat{\cS}_{I}|}{|\hat{\cS}_j|}.
\end{align}
Further, if $(1-\epsilon)c^* < \sum_{j=1}^{k} \frac{|\hat{\cS}_{I}|}{|\hat{\cS}_j|} < (1+\epsilon)c^*$ for any $\epsilon \in (0,1)$, then $|\frac{\bar{\lambda}}{\bar{\kappa}} - 1| < \epsilon$.
\end{prop}


\subsection{Power analysis}
We also establish a lower bound for the Power based on $\left\{ {\rm Power}_i\right\}$, $i=1,\dots,k$ as well as the power produced by the Union bound, before which we introduce the basic lemma to establish the connection between overall true positive proportion (TPP) with machine-wise $\TPP_i$, $i=1,\dots,k$.
\begin{lem}\label{lem:shrink}
Based on the ADAGES algorithm, we obtain
\begin{align}
\TPP \geq \frac{1}{k} \sum_{i=1}^{k} \TPP_i - \frac{c^*}{k} \frac{| \cS^c \cap \hat{\cS}_{(c^*)} |}{| \cS |}.
\end{align}
\end{lem}
The second term $\frac{c^*}{k} \frac{| \cS^c \cap \hat{\cS}_{(c^*)} |}{| \cS |}$ acts as the term of ``power shrinkage'' and can be connected with ${\rm FDP}$ in the form:
\begin{align}
{\rm Power~shrinkage} = \frac{c^*|\hat{\cS}_{(c^*)} |}{k| \cS |}{\rm FDP},
\end{align}
which involves a tradeoff between $c^*$ and $|\hat{\cS}_{(c^*)} |$.
Therefore, with proper restriction on $|\hat{\cS}_{(c^*)} |$, i.e. a proper choice of $c^*$, we can simultaneously control $\FDR$ and the power shrinkage term, which is shown in theorem~\ref{thm2}.
\begin{thm}\label{thm2}
Denote ${\rm Power}_i$ as the selection of for the $i$th machine. Suppose there exists constant $\gamma \in (0,1/2)$ such that $|\hat{\cS}_{(c^*)} | \leq (1+\gamma)|\cS|$ and $c^* \leq  k/2$. If the overall $\FDR$ is controlled at level $q \in (0,1)$,
then for a constant $\alpha \leq 3/4$, we have
\begin{align}
{\rm Power} \geq \frac{1}{k} \sum_{i=1}^{k} {\rm Power}_i - \alpha q.
\end{align}
\end{thm}

It is noticeable that power produced by the Union bound is the maximum power one aggregation method can achieve. Denote ${\rm diff} = |(\hat{\cS}_{U}\cap \cS)\backslash(\hat{\cS}_{U}\cap \cS)| = |(\hat{\cS}_{U}\cap \cS)| - |(\hat{\cS}\cap \cS)|$, with which we obtain the following theorem.
\begin{thm}\label{thm3}
Suppose we have a uniform lower bound for ${\rm Power}_i,i\in [k]$ that $\PP(j \in \cS, j \in \hat{\cS}_i) \geq \eta_{n,d}$ for $i\in [k],j \in[d]$. If we further have $c^* \leq k/2$, then $\exists \xi \leq 2$ such that
\begin{align}
    \EE[{\rm diff}] \leq \xi (1-\eta_{n,d})|\cS|.
\end{align}
Further, if the selection method has the property that $\eta_{n,d} \rightarrow 1$ as $n,d \rightarrow \infty$, we have $|{\rm Power} - {\rm Power}_{U}| \rightarrow 0$ as $n,d \rightarrow \infty$.
\end{thm}

\section{Numerical simulation}\label{simu}
In this section, we study the performance of our adaptive aggregation method by comparisons with the empirical Union, Intersection and median-aggregation rules in simulations. We also compare with the performance of the aggregation method in \citep{xie2019aggregated}, which is a modified version of the Union rule. In numerical simulations, we use model-X knockoffs with second-order construction for each machine which produces exact $\FDR$ control, so the method can be named as ``model-X knockoffs + ADAGES'' to illustrate the procedure. In this case, we are also interested in the comparison between our algorithm-free ADAGES and the knockoff-based aggregation method AKO in \citep{nguyen2020aggregation}. We consider the AKO with BY step-up with theoretical guarantee and use $\gamma = 0.3$ that is adopted in \citep{nguyen2020aggregation}. In experiments, ADAGES refers to our adaptive method with $c^* = {\rm argmin} \left\{\eta_c: 1 \leq c \leq c_0\right\}$ while ${\rm ADAGES}_m$ is the modified method with threshold $\widetilde{c} = {\rm argmin}_{1 \leq c \leq c_0} c |\hat{\cS}_{(c)}|$.

A simple linear model is adopted for feature selection:
\begin{align}
\mathbf{y} = \mathbf{X} \bfm{\beta} + \mathbf{\epsilon},
\end{align}
where $\mathbf{X} \in \RR^{n \times d} \sim \cN(\mathbf{0},\bSigma)$ is the design matrix, where $\bSigma \in \RR^{d \times d}$ and $\bSigma_{ls} = \rho^{|l-s|}$ for all $l,s \in [d]$. $\mathbf{y} \in \RR^{n}$ is the vector of $n$ responses and elements in the noise vector $\mathbf{\epsilon}$ are drawn {\it i.i.d.} from standard Gaussian distribution. Feature importance is revealed in $\mathbf{\beta}$ and $\cS = \left\{j \in \left[d\right]: \beta_j \neq 0\right\}$. 

Comparisons are conducted in the following two aspects, in which the repetition number is $r=100$ and $\rho = 0.25$. We use the criteria of averaged FDP and averaged power as the sample-versions of FDR and power respectively.
\begin{figure*}[t]
  \centering
  \includegraphics[width=\textwidth]{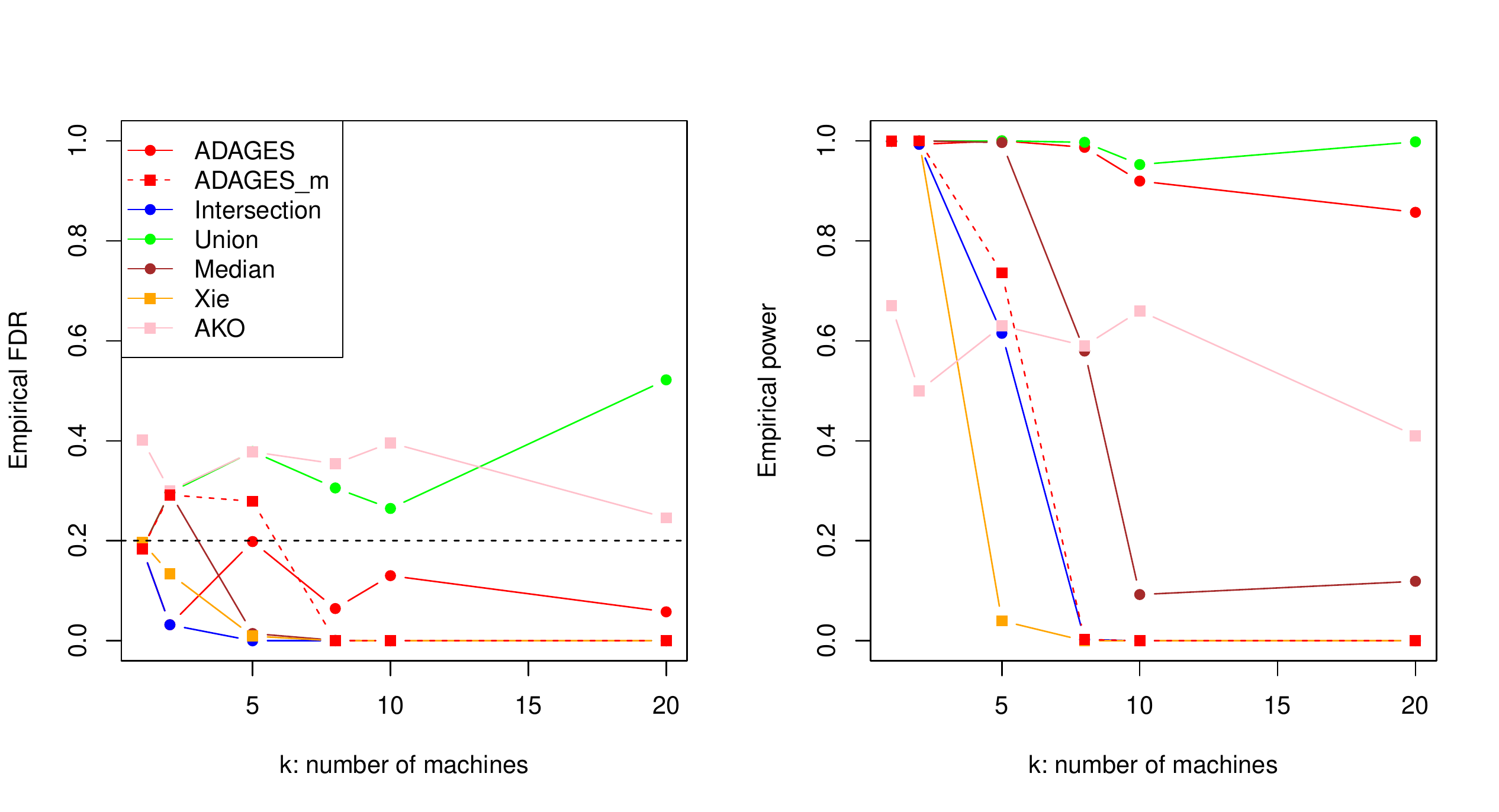}
  \caption{Left: FDP against number of machines $k$; Right: empirical power against number of machines $k$ ($k\in \left\{1,2,5,8,10,20\right\}$ with $n=1000$, $d=50$ and $s=|\cS | = 20$ over $100$ trials).}
  \label{fig:num_k}
\end{figure*}
\subsection{Varying the number of machines $k$}
Since the number of machines is a vital factor in the context of distributed learning, in the first experiment, we vary $k$ among $\left\{1,2,5,8,10,20\right\}$ with $n=1000$, $d=50$ and $s=|\cS | = 20$ fixed. Here nonzero elements in true $\mathbf{\beta}$ is drawn {\it i.i.d.} and uniformly from $\left\{\pm 2\right\}$.

From Figure \ref{fig:num_k}, we can see that ADAGES obtains a desirable tradeoff between the averaged FDP and power. As an adaptive aggregation method, ADAGES controls FDP exactly under $q=0.2$ while achieves power nearly as good as that of the Union rule, which meets the goal of power maintenance for controlled feature selection. For the three empirical methods, although the Union rule maintains power at the highest level, it produces FDP exceeding the pre-defined level $q=0.2$; the Intersection rule has conservative control of FDP but results in a serious loss of power in feature selection while the power loss of median-aggregation occurs earlier than ADAGES.

As an improvement for the Union rule on $\FDR$ control, the method in \citep{xie2019aggregated} obtains comparable FDP with the Intersection rule; but since the pre-defined level for each machine becomes $q_i = q/k$, this method will sacrifice power as shown in Figure \ref{fig:num_k} and is thus limited in application. On the other hand, in this case without ultra-high dimension or strict sparsity, the AKO that transforms more informative ``p-values'' in aggregation is capable of controlling the averaged FDP around the level $\kappa q$ where $\kappa \leq 3.24$ is given in \citep{nguyen2020aggregation}; power of AKO is lower than other algorithm-free methods when $k < 10$, but remains stable as $k$ increases. 

However, the modified ADAGES with $\widetilde{c} = {\rm argmin}_{1 \leq c \leq c_0} c |\hat{\cS}_{(c)}|$ does not produce higher power in experiments since the power shrinkage term indicates the tradeoff between FDP and $c |\hat{\cS}_{(c)}|$. Here, FDP is also a function of $c$ which ought not to be ignored in the choice of $\widetilde{c}$.

\begin{figure*}[t]
  \centering
  \includegraphics[width=\textwidth]{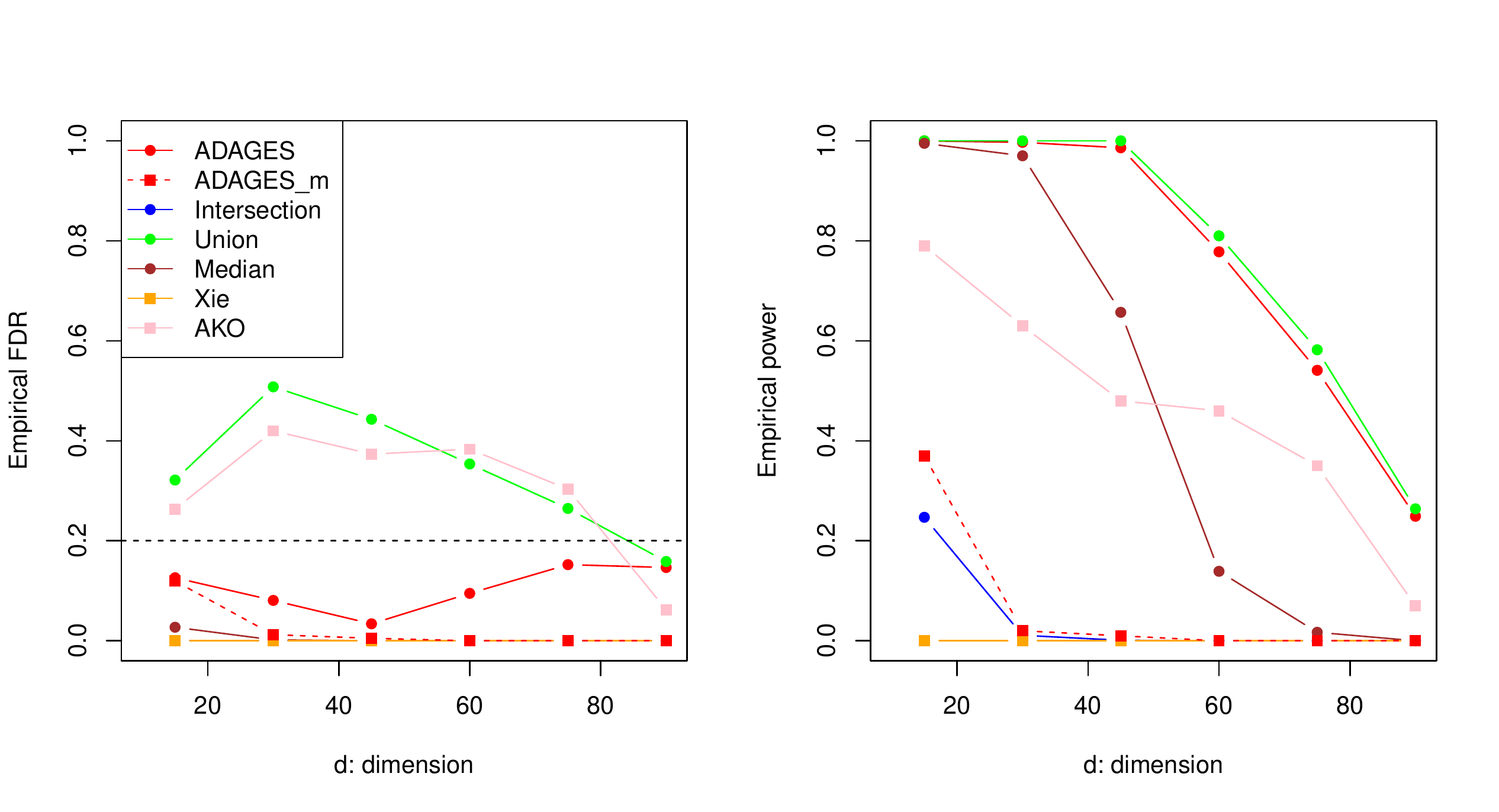}
  \caption{Left: FDP against dimension $d$; Right: empirical power against dimension $d$ ($d\in\left\{15,30,45,60,75,90\right\}$ with $n=1000$, $k=10$ and $s=|\cS | = 10$ over $100$ trials).}
  \label{fig:dim}
\end{figure*}
\subsection{Varying dimension $d$}
In the second experiment, we vary dimension $d$ in the set \\$\left\{15,30,45,60,75,90\right\}$ while fix model parameters as $n=1000$, $k=10$ and $|\cS | = 10$. True signal $\beta_j$ is generated in the same way mentioned above.

In Figure \ref{fig:dim}, both ADAGES, median-aggregation and the Intersection rule have exact FDP control under $q=0.2$, but the Union rule suffers from ``uncensored'' aggregation and cannot control the overall FDP. Partially dependent on the property of the feature selection method adopted for each machine, the power goes down as $d$ increases. But it is noticeable that the Union rule can always achieve the highest power after aggregation and ADAGES shows comparable performance due to the use of an adaptive threshold based on $\eta_c$ on an interval with an upper bound. 

In addition, the aggregation method in \citep{xie2019aggregated} tends to make null discovery that is $|\hat{\cS}| = 0$ which naturally control $\FDR$ at 0 but also have no power. Similar to our findings with varying $k$, the AKO performs better than empirical aggregation methods as $d$ increases, especially in power; but ADAGES shows better performance in both averaged FDP and empirical power.

\section{Discussion}
In this paper, we present an adaptive aggregation method called ADAGES for distributed false discovery rate control. Our method utilizes selected subsets from all machines to determine the aggregation threshold and shows better performance in the tradeoff of $\FDR$ control and power maintenance compared with empirical aggregation methods. The ADAGES is algorithm-free, which means it can be applied to any machine-wise feature selection method, and is thus more flexible than aggregation rules based on specific statistics produced by each machine-wise method.
It is motivating to further study the modified method based on the power shrinkage term, which has theoretical intuition for power maintenance and requires a good estimation of overall FDP.

Besides, as potential extensions, we can adopt this adaptive method with stability in other statistical aspects in distributed learning. Selected subsets are binary vectors consisting of limited but private information and we can further take communication constraints and privacy into consideration, which are left for our future work. More importantly, there is a tradeoff between information communication and selection power, thus it is meaningful to study aggregation methods with machines transferring encoded but more informative statistics.

As the distributed pattern becomes more common in the statistical community, to promote inter-institutional collaboration, efficient aggregation methods are necessary for distributed computing as well as privacy protection. With the idea of adaptive aggregation, collaboration can adapt to specific scenarios while each institution simply needs to focus on its specific statistical problem, which greatly contributes to the new collaboration mode in data science. However, another direction for future research is to relax the independence assumption among institutions in the learning procedure and to study the influence of inter-institutional dependence in the statistical context.

Implementation of ADAGES with R is available and raw codes can be accessed on \url{https://github.com/yugjerry/ADAGES/blob/master/code_ADAGES.R}. Technical proofs are presented in the following sections.

\appendix
\section{Technical proofs}\label{proof}
In this section, we present the proofs for the main theorems and propositions in this paper.

\noindent \textbf{Proof for Theorem~\ref{thm1}.}
With $\FDR_i = \EE\left[\frac{|\hat{\cS}_i \cap \cS^{c}|}{|\hat{\cS}_i|}\right] \leq q$, observe the overall $\FDR$:
\begin{align}
\FDR_{\left(c^*\right)} &= \EE\left[\frac{|\hat{\cS}_{(c^*)} \cap \cS^{c}|}{|\hat{\cS}_{(c^*)}|}\right] 
\nonumber\\&= \EE\left\{\EE\left[\frac{|\hat{\cS}_{(c^*)} \cap \cS^{c}|}{|\hat{\cS}_{(c^*)}|}\Big|\left(\hat{\cS}_1,\dots,\hat{\cS}_k\right)\right]\right\}.
\end{align}
First, we have 
$c^* |\hat{\cS}_{(c^*)}| \leq \sum_{j:m_j \geq c^*} m_j \leq \sum_{j=1}^{d} m_j = \sum_{i=1}^{k} |\hat{\cS}_i|$,
and similarly for features $j \in \cS^{c}$,
\begin{align}
c^* |\hat{\cS}_{(c^*)} \cap \cS^{c}| \leq \sum_{j \in \cS^{c}: m_j \geq c^*} m_j \leq \sum_{j \in \cS^{c}} m_j = \sum_{i=1}^{k} |\hat{\cS}_i \cap \cS^{c}|.
\end{align}
Therefore, the overall $\FDR$ can be linked to the machine-wise $\FDR$'s as 
\begin{align}
\FDR_{\left(c^*\right)} &= \EE\left\{\EE\left[\frac{|\hat{\cS}_{(c^*)} \cap \cS^{c}|}{|\hat{\cS}_{(c^*)}|}\Big|\left(\hat{\cS}_1,\dots,\hat{\cS}_k\right)\right]\right\}
\nonumber\\&\leq 
 \EE\left\{\frac{1}{c^*}\sum_{i=1}^{k} \EE\left[\frac{|\hat{\cS}_i \cap \cS^{c}|}{|\hat{\cS}_{(c^*)}|}\Big|\left(\hat{\cS}_1,\dots,\hat{\cS}_k\right)\right]\right\}.
\end{align}
In addition, by definition of $c^*$ in the theorem: $|\hat{\cS}_{(c^*)}| \geq \frac{1}{k} \sum_{i=1}^{k} |\hat{\cS}_i| \geq \frac{k}{\sum_{i=1}^{k} 1/|\hat{\cS}_i|}$, we then obtain
\begin{align}
\FDR_{\left(c^*\right)} & \leq \EE\left\{\frac{1}{c^*}\sum_{i=1}^{k} \EE\left[\frac{|\hat{\cS}_i \cap \cS^{c}|}{|\hat{\cS}_{(c^*)}|}\Big|\left(\hat{\cS}_1,\dots,\hat{\cS}_k\right)\right]\right\}
\nonumber\\
&\leq \EE\left\{\frac{1}{k c^*}\sum_{j=1}^{k} \sum_{i=1}^{k} \EE\left[\frac{|\hat{\cS}_i \cap \cS^{c}|}{|\hat{\cS}_j|}\Big|\left(\hat{\cS}_1,\dots,\hat{\cS}_k\right)\right]\right\}\nonumber\\
& = \EE\left\{\frac{1}{k c^*}\sum_{1 \leq i,j \leq k} \EE\left[\frac{|\hat{\cS}_i \cap \cS^{c}|}{|\hat{\cS}_i|} \cdot \frac{|\hat{\cS}_i|}{|\hat{\cS}_j|}\Big|\{\hat{\cS}_l\}_{l=1}^{k}\right]\right\}\nonumber\\
& = \EE\left\{\frac{1}{k c^*}\sum_{i=1}^{k} \sum_{j=1}^{k}\EE\left[{\rm FDP}_i\frac{|\hat{\cS}_i|}{|\hat{\cS}_j|} \Big|\left(\hat{\cS}_1,\dots,\hat{\cS}_k\right)\right]\right\} 
\nonumber\\
& \leq \sum_{i = 1}^{k} \frac{1}{kc^*} {\rm FDP}_i \cdot \left(\max_{1\leq i \leq k}|\hat{\cS}_i| \sum_{j=1}^{k} \frac{1}{|\hat{\cS}_j|}\right)\nonumber\\ 
& \leq \sum_{i = 1}^{k} \frac{1}{kc^*} {\rm FDP}_i \cdot \lambda c^* \leq \lambda q.
\end{align}
Here, $\lambda$ is a bound that 
\begin{align}
    \lambda \geq \max_{1\leq i \leq k}\frac{|\hat{\cS}_i|}{c^*} \sum_{j=1}^{k} \frac{1}{|\hat{\cS}_j|}.
\end{align}
$\ep$
\\
\\
\textbf{Proof for theorem~\ref{thm2}.}
We consider the expected number of true discoveries $\EE| \hat{\cS} \cap \cS |$ and denote 
${\rm TPP}_i = \frac{|\hat{\cS}_i \cap \cS|}{| \cS |},~i \in \left[k\right]$ and $\TPP = \frac{|\hat{\cS}_{(c^*)} \cap \cS|}{| \cS |}$. Then we have
\begin{align}
|\cS| \sum_{i=1}^{k} \TPP_i &= \sum_{i=1}^{k} | \hat{\cS}_i \cap \cS|
 \nonumber\\&= \sum_{i=1}^{k} \sum_{j \in \cS} \bfm{1}_{\left\{j \in \hat{\cS}_i\right\}} = \sum_{j \in \cS} m_j 
 \nonumber\\&= \sum_{j \in \cS \cap \hat{\cS}_{(c^*)}} m_j + \sum_{j \in \cS^c \cap \hat{\cS}_{(c^*)}} m_j 
 \nonumber\\&\leq k| \cS \cap \hat{\cS}_{(c^*)} | + c^* | \cS^c \cap \hat{\cS}_{(c^*)} |,
\end{align}
which is equivalent to 
\begin{align}
\TPP \geq \frac{1}{k} \left(\sum_{i=1}^{k} \TPP_i - c^* \frac{| \cS^c \cap \hat{\cS}_{(c^*)} |}{| \cS |} \right).
\end{align}
Based on the assumption with an upper bound on $\hat{\cS}_{(c^*)}$ with $\gamma \in (0,1/2)$, 
\begin{align}
|\hat{\cS}_{(c^*)}|\leq (1+\gamma) |\cS|,
\end{align}
we take expectation for the inequality and then obtain
\begin{align}
{\rm Power} \geq \frac{1}{k} \sum_{i=1}^{k} {\rm Power}_i - \alpha q,
\end{align}
where $\alpha = \frac{c^*}{k}(1+\gamma) < \frac{3}{4}$.
$\ep$
\\
\\
\textbf{Proof for theorem~\ref{thm3}.}
We can write explicitly that 
\begin{align}
    {\rm diff} = \sum_{j \in \cS} \bfm{1}_{\left\{0 < m_j < c^*\right\}}.
\end{align}
Then, for positive $m_j$, $\EE[{\rm diff}] = \sum_{j \in \cS} \PP(m_j < c^*)$ with 
\begin{align}
    \PP(m_j < c^*, j \in \cS) &\leq \frac{k - \EE[m_j|j \in \cS]}{k - c^*} \nonumber\\&= \frac{k - \EE[\sum_{i=1}^{k} \bfm{1}_{\{j \in \hat{\cS}_i\}}|j \in \cS]}{k - c^*} \nonumber\\&\leq \frac{k(1-\eta_{n,d})}{k - c^*} \nonumber\\
    &\leq \xi (1-\eta_{n,d}),
\end{align}
where $c^* \leq k/2$ by definition and thus $\xi \leq 2$.
$\ep$
\\
\\
\noindent\textbf{Proof for proposition~\ref{union}.}
With the Union rule, $\hat{\cS}_{U} = \bigcup_{i=1}^{k} \hat{\cS}_i$ and thus $\hat{\cS}_i \subset \hat{\cS}_{U}$ for all $i=1,\dots,k$. Since
$|\bigcup_{i=1}^{k} A_i| \leq \sum_{i=1}^{k} |A_i|$,
we apply this fact to $\hat{\cS}_{U} \cap \cS^c = \bigcup_{i=1}^{k} (\hat{\cS}_i \cap \cS^c)$ and consider the overall FDR:
\begin{align}
\FDR &= \EE\left[\frac{|\hat{\cS}_{U} \cap \cS^c|}{|\hat{\cS}_{U}|}\right] \leq \EE\left[\sum_{i=1}^{k} \frac{|\hat{\cS}_{i} \cap \cS^c|}{|\hat{\cS}_{U}|}\right] \nonumber\\&\leq \sum_{i=1}^{k} \EE\left[\frac{|\hat{\cS}_{i} \cap \cS^c|}{|\hat{\cS}_{i}|} \right] \leq \sum_{i=1}^{k} \FDR_i \nonumber\\&\leq \sum_{i=1}^{k} q_i.
\end{align}
$\ep$
\\
\\
\noindent\textbf{Proof for proposition~\ref{intersec}.}
With $\hat{\cS}_{I} = \bigcap_{i=1}^{k} \hat{\cS}_i$, we have $m_j = \sum_{i=1}^{k} \bfm{1}_{\left\{j \in \hat{\cS}_{i}\right\}} = k$ for $j \in \hat{\cS}$. Therefore, we have 
\begin{align}
k|\hat{\cS}_{I} \cap \cS^c| &= \sum_{j \in \hat{\cS}_{I} \cap \cS^c} m_j \leq \sum_{j \in \cS^c} m_j \nonumber\\&= \sum_{j \in \cS^c} \sum_{i=1}^{k} \bfm{1}_{\left\{j \in \hat{\cS}_{i}\right\}} =  \sum_{i=1}^{k} |\hat{\cS}_{i} \cap \cS^c|
\end{align}

We then consider the overall FDR,
\begin{align}
\FDR &= \EE\left[\frac{|\hat{\cS}_{I} \cap \cS^c|}{|\hat{\cS}_{I}|}\right] \leq \EE\left[\frac{1}{k}\frac{|\hat{\cS}_{i} \cap \cS^c|}{|\hat{\cS}_{I}|}\right]\nonumber\\
&\leq \EE\left[\frac{1}{k}\frac{|\hat{\cS}_{i} \cap \cS^c|}{|\hat{\cS}_{i}|} \cdot \frac{|\hat{\cS}_{i}}{|\hat{\cS}_{I}|}\right] \nonumber\\&\leq \frac{\kappa}{k} \sum_{i=1}^{k} \EE\left[\frac{|\hat{\cS}_{i} \cap \cS^c|}{|\hat{\cS}_{i}|}\right]\leq \kappa q.
\end{align}
Here $\kappa \geq 1$ is a constant such that
$\max_{i \in \left[k\right]} \frac{|\hat{\cS}_i|}{|\hat{\cS}_{I}|} \leq \kappa$.
$\ep$

\section{Illustration of the aggregation process}
In this part, results in four cases are provided to illustrate the connection between overall FDR/power and the machine-wise ones. The $k$ grey bars in each plot are the $ \FDR_i$s or ${\rm Power }_i$s for $k$ machines.

From the four cases together with the simulation results in our paper, we can see that ADAGES has a better tradeoff than other methods (the Union rule, the Intersection rule, median aggregation and method in \citep{xie2019aggregated}). For FDR, all methods except the Union rule produce the exact control whenever machine-wise FDR is controlled at the pre-defined level. The Union rule, however, as is shown in proposition~\ref{union}, is only able to control FDR at a higher level. When $k$ or the dimension $d$ is large, strict aggregation methods will cause the power loss, such as the results of the Intersection rule and method in \citep{xie2019aggregated}. We should note that ``strict'' refers to strict pre-defined levels for each machine as well as strict aggregation rules. As is shown in the results, ADAGES produces power very close to that of the Union rule, which is the highest power an aggregation method can achieve.

\begin{figure}[H]
  \centering
  \includegraphics[width=0.8\textwidth]{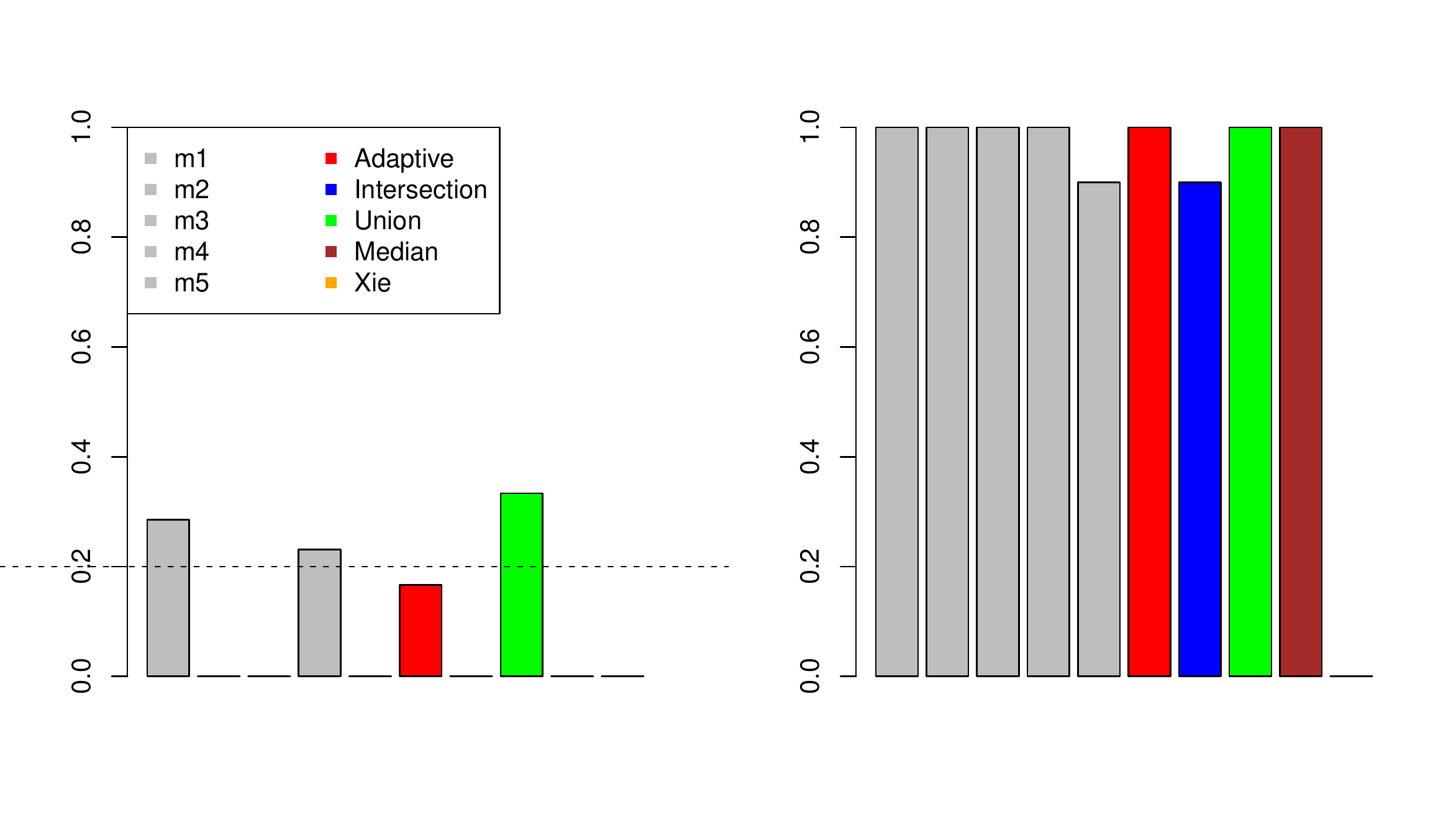}
  \caption{Representation of the aggregation process: barplot of machine-wise FDR(left)/power(right) and aggregation results under different rules ($q=0.2,k=5,d=20,n=1000,n_i=200$).}
  \label{fig:p1}
\end{figure}

\begin{figure}[H]
  \centering
  \includegraphics[width=0.8\textwidth]{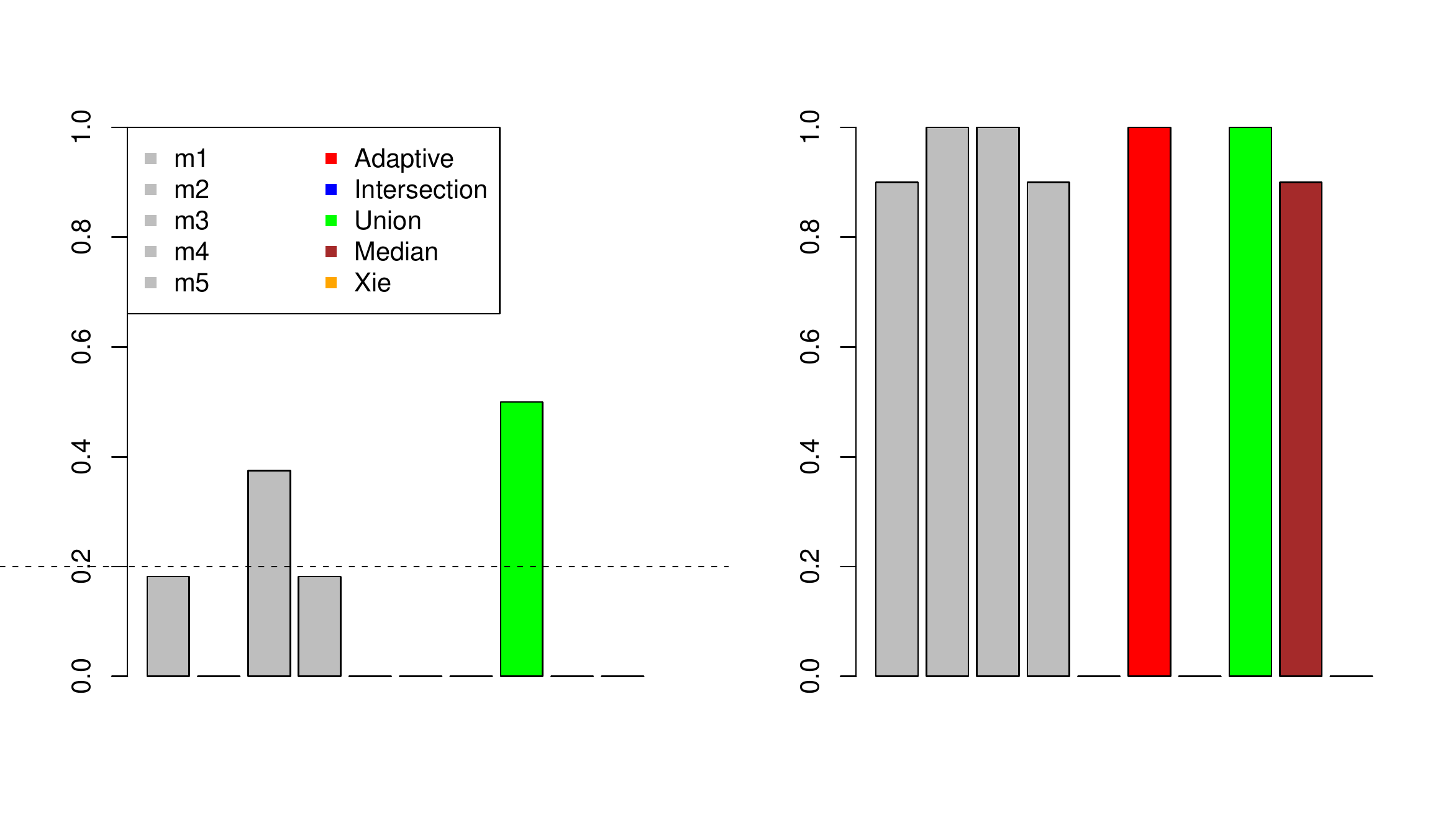}
  \caption{Representation of the aggregation process: barplot of machine-wise FDR(left)/power(right) and aggregation results under different rules ($q=0.2,k=5,d=80,n=1000,n_i=200$).}
  \label{fig:p2}
\end{figure}

\begin{figure}[H]
  \centering
  \includegraphics[width=0.8\textwidth]{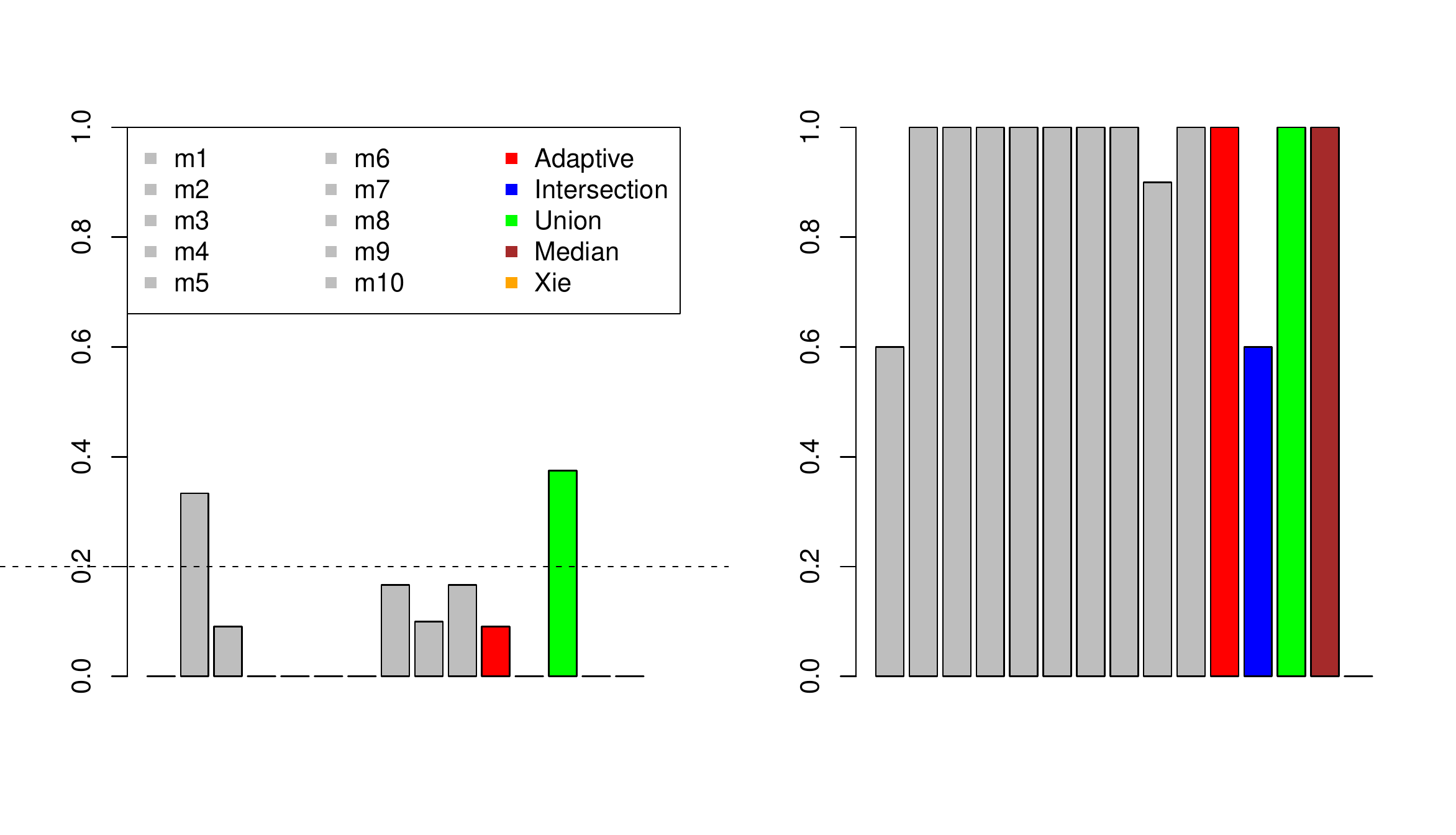}
  \caption{Representation of the aggregation process: barplot of machine-wise FDR(left)/power(right) and aggregation results under different rules ($q=0.2,k=10,d=20,n=1000,n_i=100$).}
  \label{fig:p3}
\end{figure}

\begin{figure}[H]
  \centering
  \includegraphics[width=0.8\textwidth]{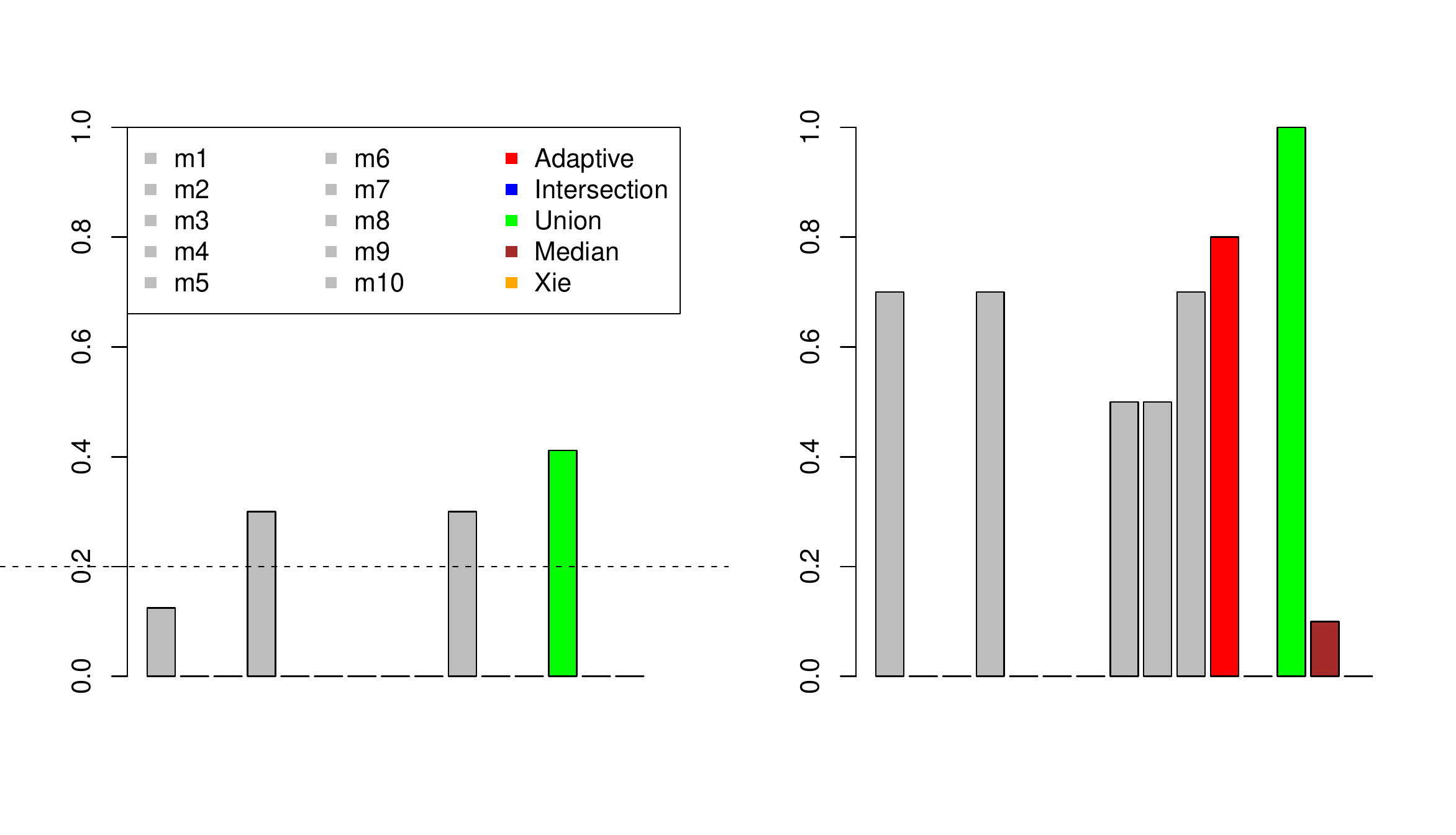}
  \caption{Representation of the aggregation process: barplot of machine-wise FDR(left)/power(right) and aggregation results under different rules ($q=0.2,k=10,d=80,n=1000,n_i=100$).}
  \label{fig:p4}
\end{figure}


\bibliographystyle{chicago}
\bibliography{ygbib}

\end{document}